\documentclass{pasj00}

\begin{document}
\SetRunningHead{MIN et al.}{Parallax Measurement toward R Aquarii}
\Received{}
\Accepted{2013/11/23}
\Published{}

\title{Accurate Parallax Measurement toward the Symbiotic Star R Aquarii}

\author{%
   Cheulhong \textsc{Min},\altaffilmark{1,2}
   Naoko \textsc{Matsumoto},\altaffilmark{2}
   Mi Kyoung \textsc{Kim},\altaffilmark{3}
   Tomoya \textsc{Hirota},\altaffilmark{1,2}
   
   Katsunori M. \textsc{Shibata},\altaffilmark{1,2}
   Se-Hyung \textsc{Cho},\altaffilmark{3}
   Makoto \textsc{Shizugami},\altaffilmark{2}
   Mareki \textsc{Honma}\altaffilmark{1,2}}

\altaffiltext{1}{Department of Astronomical Sciences, The Graduate University for Advanced Studies, 2-21-1, Osawa, Mitaka, Tokyo 181-8588}
\altaffiltext{2}{Mizusawa VLBI observatory, National Astronomical Observatory of Japan, 2-21-1, Osawa, Mitaka, Tokyo 181-8588}
\altaffiltext{3}{Korea Astronomy and Space Science Institute, 776 Daedukdaero (61-1 Whaam), Yuseong, Daejeon 305-348 KOREA}
\email{cheulhong.min@nao.ac.jp}


\maketitle

\begin{abstract}
Multi-epoch phase-referencing VLBI (Very Long Baseline Interferometry) observations with VERA (VLBI Exploration of Radio Astrometry) were performed for the symbiotic star R Aquarii (R Aqr) from September 2005 to Oct 2006. Tracing one of the $v=2$, $J=1-0$ SiO maser spots, we measured an annual parallax of $\pi = 4.59\pm0.24$ mas, corresponding to a distance of $218_{-11}^{+12}$ pc. Our result is consistent with earlier distance measurements, but yields the highest accuracy of about 5\% level. Applying our distance, we derived an absolute K-band magnitude of $M_{\mathrm{K}} = -7.71 \pm 0.11$, which is consistent with the recent Period-Luminosity relation by VLBI parallax measurements for 5 OH-Mira variables. In addition, the expansion age of an inner nebulae around R Aqr is found to be about  240 years, corresponds to about the year 1773. \\ 

\end{abstract}

\newpage
\section{Introduction}
The class of Symbiotic star was first introduced by Merill (1958), who presented the characteristic spectrum showing a combination of the giant star continuum with molecular absorption features and ionized emission lines. These stars are nowadays generally understood as interacting binary system comprising of a cool late-type star, which is a red giant (RG) or asymptotic giant branch (AGB) star, and a hot compact companion, usually a white dwarf or in a few cases of a low-mass main-sequence star. Phenomena caused by interaction between the components in symbiotic stars, such as accretions of the stellar material around the hot component with nova-like thermonuclear outburst, mass-loss from a giant star with colliding wind process, formation of bipolar photo-ionized planetary nebulae, and collimation of a jet-like feature, give unique astrophysical laboratories to investigate the evolution of binary systems (Kenyon 1986).

R Aquarii (R Aqr) is one of the most studied symbiotic stars composed of a Mira, a long period variable, with a pulsation period of about 387 days and a white dwarf companion as well as an ionized nebulae around the system. Unusually, R Aqr exhibits an astronomical jet feature that is probably powered by an accretion disk around a white dwarf companion. This jet has been extensively observed in optical, radio, UV, and X-ray wavelengths and is known to extend up to about 2500 AU in a NE-SW direction with a shock speed of about 235 - 285 km s$^{-1}$ (Nichols \& Slavin 2009).

The large-scaled, extended inner and outer nebulae are also known in R Aqr. Solf \& Ulrich (1985) carried out an extensive study of the nebulae using optical spectroscopy observations, and revealed that both nebulae had the same geometric structure resembling a bipolar, hourglass-like shape extending about two arcmin in an East-West direction for the outer nebula with an equatorial velocity of 55 km s$^{-1}$ and about one arcmin in a North-South direction for the inner nebula with an equatorial velocity of 32 km s$^{-1}$, respectively. 

An orbital period of 44 years for R Aqr was proposed by Willson et al. (1981), who interpreted that the depressions of the optical light curve from 1928 to 1934 and from 1974 to 1978 could be attributed to the obscuration of the Mira variable by an extended cloud of dust. A wide-range of radial velocity data were collected by McIntosh \& Rustan (2007) in visual, Near-IR, and radio wavelengths from 1940. They extracted the orbital period of 34.6 years with an eccentricity of 0.52, and a projected semi-major axis of 3.5 AU. Subsequently, Gromadzki \& Miko\l{}ajewska (2009) also extracted the orbital period of 43.6 years with an eccentricity of 0.25 modifying the radial velocity data collected by McIntosh \& Rustan (2007) complemented by additional radial velocity data. They obtained a consistent orbital period with Willson et al. (1981). 

In addition, R Aqr is one of three symbiotic stars that have circumstellar masers associated with Mira variables. Since the first detection of an SiO maser toward R Aqr by L\'{e}pine et al. (1978), there have been several single-dish observational studies of 43/86/128 GHz SiO masers (Zuckerman 1979; Cohen \& Ghigo 1980; Spencer et al. 1981; Mart\'{i}nez et al. 1988; Jewell et al. 1991; Schwarz et al. 1995; Cho et al. 1996; Alcolea et al. 1999; Pardo et al. 2004; Kang et al. 2006) as well as 22/321 GHz H$_{2}$O maser (Seaquist et al. 1995; Ivison et al 1994, 1998). Moreover, high-resolution VLBI observations were also performed with VLBA presenting a ring-like structure of the 43 GHz SiO maser with an approximated diameter of 30 mas (Boboltz et al. 1997; Hollis et al. 2001; Cotton et al. 2004, 2006). 

VLBI monitor observations of SiO masers toward R Aqr were carried out since 2004 by Kamohara et al. (2010) with VERA. They observed both $v=1$ and $v=2$, $J=1-0$ SiO maser transitions, and confirmed that both maser emissions appear in similar regions. However, each transition of maser spots were not exactly  coincident within the spatial and spectral resolutions, so that new theoretical studies were required for looking into the finer details of both maser distributions in the inner shells around the central AGB star. They also estimated an annual parallax of $4.7 \pm 0.8$ mas assuming that the position of the star is coincident with the center of circular fitting of both SiO maser distributions, because of the short lifetime of individual SiO maser spots. 

A typical lifetime of an SiO maser feature is on the order of one hundred days, but a few cases of the SiO maser features persist for a period of one year. The best example is high-resolution VLBA monitoring observations for the SiO maser in Mira variable of TX Cam (Diamond \& Kemball 1999, 2003; Gonidakis et al. 2010). According to the results by Diamond \& Kemball (2003) and Gonidakis et al. (2010), the lifetime of the SiO maser components around the Mira variable was suggested to be between 150 and 200 days, but sometimes long-lived maser components persisted over 350 days. Moreover, McIntosh \& Bonde (2010) reported that the average lifetime of the SiO maser features was 171 days, but also reported the existence of long-lived SiO maser features over 171 days. 

The VERA array of the National Astronomical Observatory of Japan (NAOJ) is a Japanese VLBI array aimed at obtaining 10 micro-arcsecond-level accuracy of parallaxes and proper motions of H$_{2}$O, SiO, and CH$_{3}$OH maser sources using its unique dual-beam phase-referencing technique. In past years, VERA has successfully measured parallaxes for maser sources in our galaxy including AGB stars, such as S Crt (Nakagawa et al. 2008), SY Scl (Nyu et al. 2011), and RX Boo (Kamezaki et al. 2012). Accurate measurements of distances for AGB stars would be useful for calibrating the Period-Luminosity relation precisely, which provides a fundamental basis of a distance ladder of nearby galaxies.

In this paper, we report phase-referencing observations of the SiO maser toward the symbiotic star R Aqr using VERA, and present the result of the most accurate parallax measurement.

\section{Observations and Data Reduction}
The observations of the $v=1$ and $v =2$, $J=1-0$ SiO maser transitions were performed using the four stations of VERA from September 2005 to October 2006. We reanalyzed these data which were published by  Kamohara et al. (2010). Rest frequencies of 43.122079 GHz and 42.820582 GHz were adopted for the $v=1$, $J=1-0$ and $v=2$, $J=1-0$ in this paper, respectively.

The target source of the SiO masers around R Aqr ($\alpha_{\mathrm{J2000}}$ = \timeform{23h43m49s.4616}, $\delta_{\mathrm{J2000}}$ = \timeform{-15D17'04".202}) and the phase-referencing source J2348$-$1631 ($\alpha_{\mathrm{J2000}}$ = \timeform{23h48m02s.608532},  $\delta_{\mathrm{J2000}}$ = \timeform{-16D31'12".02226}) were observed simultaneously using the VERA dual-beam system. The separation angle between the target and reference source is about 1.6 degrees. A bright continuum source, 3C454.3, was also observed every 80 minutes as a calibrator. The instrumental phase difference between the two beams was measured in real time during the observations, by correlating random signals from artificial noise sources injected into both beams at each station (Honma et al. 2008a). 

The data were recorded onto magnetic tapes at a rate of 1024 Mbps with the VERA DIR2000 recording system, providing a total bandwidth of 256 MHz with 2-bit digitization. The 256 MHz bandwidth data were divided into 16 IF channels (16 MHz each), and two of them were assigned to the $v=1$ and $v=2$, $J=1-0$ SiO maser transitions of the target source, while others were assigned to continuum spectrum of the reference and calibrator source. 
Correlation processing was carried out on the Mitaka FX correlator located in NAOJ, Mitaka. The spectral resolution of the maser lines was set to be 31.25 kHz, corresponding to a velocity resolution of 0.21 km s$^{-1}$ for observation epochs of 2005/270, 2006/207, 242, 286 (year/day of year) and 15.625 kHz, corresponding to a velocity resolution of 0.11 km s$^{-1}$ for 2005/327, 358, 2006/045, 062, 128.

All the data reductions were performed using the National Radio Astronomy Observatory (NRAO) Astronomical Image Processing System (AIPS). The amplitude and the bandpass calibrations for the target source and the reference source were made independently. For phase-referencing analysis, we calibrated the clock parameters using the calibrator, 3C454.3, and the fringe fitting was made on the position reference source, J2348-1631, to obtain residual fringe phase. These solutions were applied to the target source, R Aqr. The dual-beam phase-calibration data and the modified delay-tracking model were also applied to the target data in order to obtain accurate astrometry with VERA. Then, using AIPS task IMAGR, we obtained the $v=1$ and $v=2$ maser images which consist of 2048 pixels $\times$ 2048 pixels in size with a pixel spacing of 0.05 mas. From these images, we searched maser spots with S/N ratio larger than six as detection criterion, and measured their positions by elliptical Gaussian fitting using AIPS task SAD. Selected maser spots were considered as real if maser spots existed in more than two adjacent channel maps within the beam size.

\section{Parallax Measurement}

In order to estimate a parallax, we assumed that the motion of the SiO maser spot is the summation of a parallax motion and a linear proper motion. The linear proper motion includes an individual motion of the maser spot in the maser-emitting region and a secular motion of the central star as well as a binary motion. However, an accelerating or decelerating factor of the binary motion is negligible and considered as the linear for one year of a 44-year orbit. Then, the motion of the maser spot on the sky plane is simply described by following equations (Seidelmann 1992) : 
\begin{displaymath}
\alpha(t)\cos{\delta} = \alpha_{0} + \mu_{\alpha}^{*}(t - t_0) + \pi P_{\alpha},
\end{displaymath}
\begin{equation}
\delta(t) = \delta_{0} + \mu_{\delta}(t - t_0) + \pi P_{\delta},
\end{equation}
where ($\alpha_{0}$, $\delta_{0}$) are the initial positions at $t = t_0$, ($\mu_{\alpha}^{*}$, $\mu_{\delta}$) are the linear proper motions for the direction of right ascension and declination (Note that $\mu_{\alpha}^{*} = \mu_{\alpha}\cos\delta$), $\pi$ is the parallax, ($P_{\alpha}$, $P_{\delta}$) are the parallax factors, which are sinusoidal functions of the parallactic ellipse for a certain time in right ascension and declination due to the motion of the Earth around the Sun. Parallax factors are computed as follows :
\begin{displaymath}
P_{\alpha} = X_{\odot}\sin{\alpha_{\star}} - Y_{\odot}\cos{\alpha_{\star}},
\end{displaymath}
\begin{equation}
P_{\delta} = X_{\odot}\cos{\alpha_{\star}}\sin{\delta_{\star}} + Y_{\odot}\sin{\alpha_{\star}}\sin{\delta_{\star}} - Z_{\odot}\cos{\delta_{\star}},
\end{equation}
where ($X_{\odot},Y_{\odot},Z_{\odot}$) are the Cartesian coordinates of the Earth relative to the barycenter of the solar system at the time of observation. These values are taken by NASA Jet Propulsion Laboratory Solar System ephemeris (e.g. DE405). ($\alpha_{\star}, \delta_{\star}$) are the position of the target (the nominal position) in right ascension and declination, respectively. 

Unknown parameters, parallax ($\pi$), linear proper motions ($\mu_{\alpha}^{*}$, $\mu_{\delta}$) and initial positions ($\alpha_{0}$, $\delta_{0}$) were fitted independently with respect to right ascension and declination directions by reduced $\chi^{2}$ fittings. The astrometric position errors were set so that the reduced $\chi^{2}$ becomes unity. Then, combined fit with both directions was conducted by giving uniform weights of $\sigma_{\mathrm{RA}}$ and $\sigma_{\mathrm{Dec}}$ in right ascension and declination, respectively.

\begin{figure}
  \begin{center}
  	\includegraphics[clip=true, width=80mm]{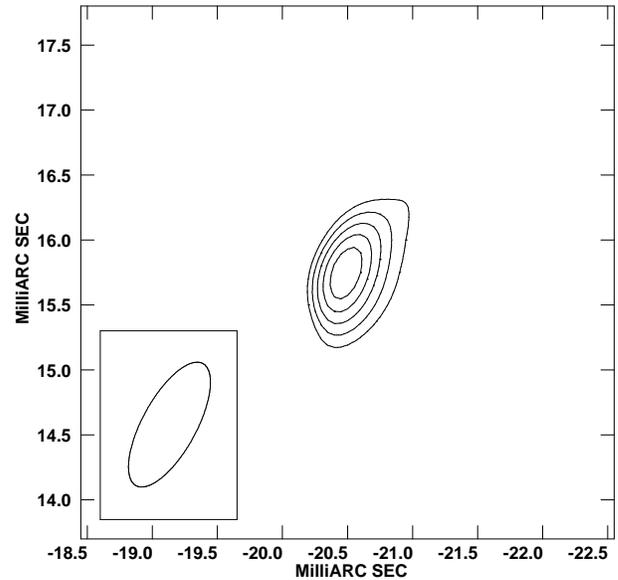}
  \end{center}
  \caption{Phase-referenced image of the $v=2$, $J=1-0$ ($V_{\mathrm{LSR}}=-20.7$ km s$^{-1}$) SiO maser spot at the first epoch on 2005/270. The $x$ and $y$-axis represent the RA and Dec offset with respect to the center coordinate of $(\alpha, \delta)_{\mathrm{J2000}} = (\timeform{23h43m 49s.4736}, \timeform{-15D17’04".3620})$. The peak intensity is 1.77 Jy beam$^{-1}$, and contour levels are set to 7, 9, 11, 13, and 15$\sigma$, where $\sigma$ is the rms noise of 0.11 Jy beam$^{-1}$. The synthesized beam is 1.07 mas $\times$ 0.42 mas (PA = $-29^{\circ}$) and is shown in the bottom-left corner of the figure}\label{fig:1}
\end{figure}

\begin{figure}
  \begin{center}
  	\includegraphics[clip=true, width=80mm]{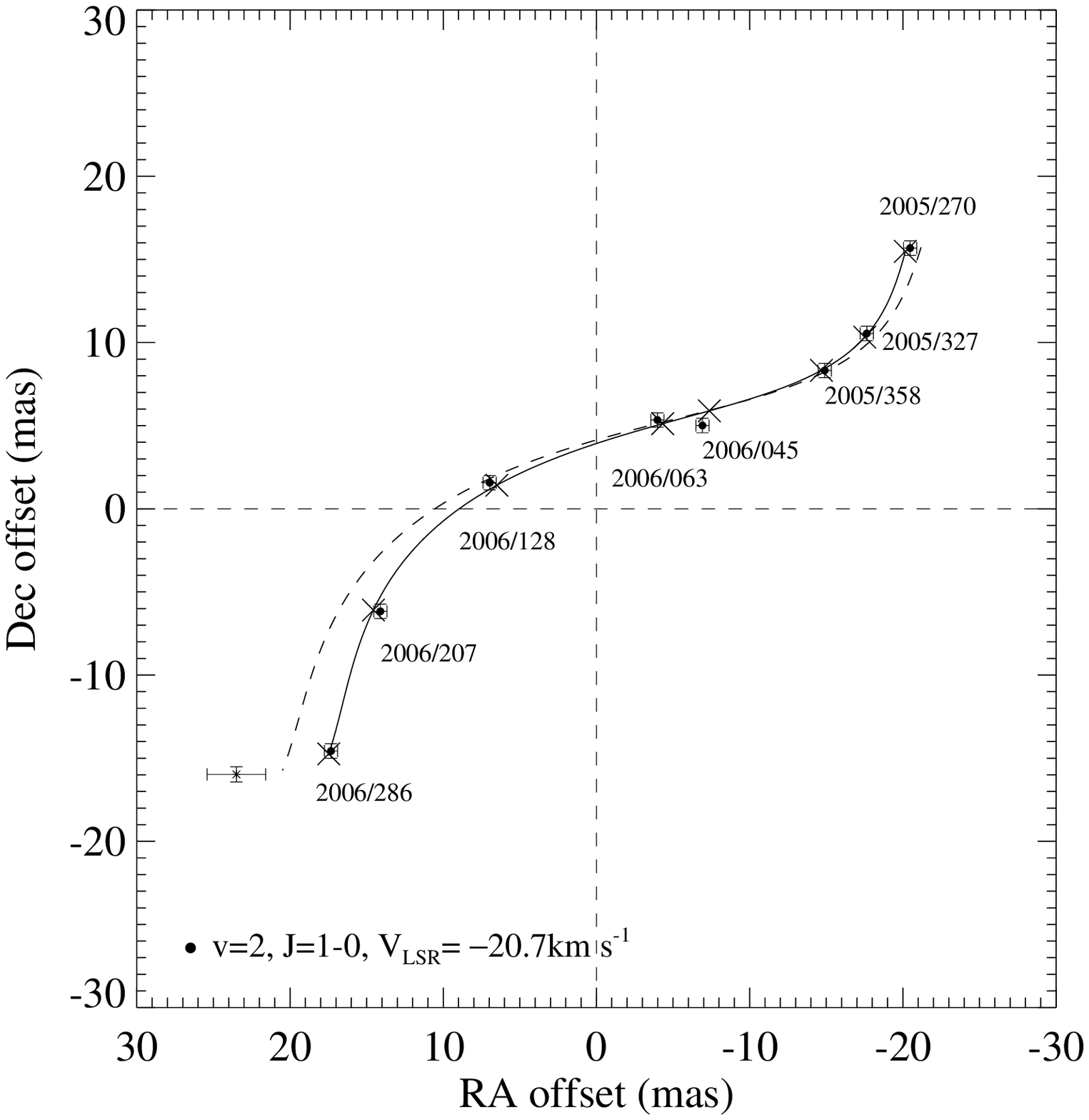}    
  \end{center}
  \caption{Measured positions of the $v=2$, $J=1-0$ ($V_{\mathrm{LSR}}=-20.7$ km s$^{-1}$) SiO maser spots. The solid line indicates the best-fitting result of parallax and proper motion. The center coordinate is $(\alpha, \delta)_{\mathrm{J2000}} = (\timeform{23h43m 49s.4736}, \timeform{-15D17’04".3620})$. The filled circles with error bars of $\sigma_{\mathrm{RA}} = 0.430$ mas and $\sigma_{\mathrm{Dec}} = 0.434$ mas in RA and Dec directions (see text) indicate the positions of traced maser spot, and cross symbols represent the predicted positions. The dashed line and the asterisk symbol with error bars indicate the misidentified case (see text). }\label{fig:2}
\end{figure}

Among all the observations, we used an SiO maser ($v=2$, $J=1-0$) spot with the $V_{\mathrm{LSR}}=-20.7$ km s$^{-1}$ to estimate parallax. Figure \ref{fig:1} shows the phase-referenced image of the target maser spot, and this spot was found to be persistent. We detected this spot in eight epochs at 2005/270, 327, 358, 2006/045, 062, 128, 207, and 286, covering a period of about one year in the same velocity channel. In Figure \ref{fig:2}, we present position variations of the traced SiO maser spot with the best-fitting parallax and proper motion results presented by solid curve. Table \ref{tab:LT1} shows the summary of the parallax and the linear proper motion results with uniform weights in right ascension and declination. By setting position errors of $\sigma_{\mathrm{RA}}=0.430$ mas and $\sigma_{\mathrm{Dec}}=0.434$ mas obtained by reduced $\chi^{2}$ fitting, we obtained the parallax of $\pi = 4.59 \pm 0.24$ mas, corresponding to a distance of $218_{-11}^{+12}$ pc. The proper motions are also obtained to be $\mu_{\alpha}^{*}=37.13 \pm 0.47$ mas yr$^{-1}$ and $\mu_{\delta}=-28.62 \pm 0.44$ mas yr$^{-1}$, respectively. 

Note that we detected another maser spot other than the target maser spot in the velocity channel with $V_{\mathrm{LSR}}=-20.7$ km s$^{-1}$ on the observation epochs of 2005/358 and 2006/286. Two maser spots were closely located within 0.51 mas on 2005/358, but separated by about 6.33 mas on 2006/286. To avoid misidentification of the target maser spot, we tried estimating parallax with each individual maser spots. For the maser spots on 2005/358, measured parallaxes for each cases presented practically the same result. On the other hand, the fitting result was significantly worse for 2006/286 when we misidentified the one of the maser spots detected on this epoch as shown in Figure \ref{fig:2}. Thus, we concluded that two maser spots on 2006/286 fall into distinct two maser features, and one of worse fitting maser spot is excluded for parallax measurement.

\begin{longtable}{cccccc}
  \caption{Parallax and linear proper motion fitting result}\label{tab:LT1}
  \hline              
	$V_{\mathrm{LSR}}$      & $\pi$ & Distance  & $\mu_{\alpha}^{*}$ & $\mu_{\delta}$  & $\sigma$ \\
	(km s$^{-1}$)  & (mas) &   (pc)    & (mas yr$^{-1}$)    & (mas yr$^{-1}$) &   (mas)  \\ \hline
 \endfirsthead
  \hline
\endhead
  \hline
\endfoot
  \hline
\endlastfoot
  \hline
		$-20.7$ & $4.65\pm0.29$ & $215_{-13}^{+15}$ & $37.08\pm0.53$ &         -       & 0.430 \\
		        & $4.38\pm0.53$ & $228_{-25}^{+32}$ &       -        & $-28.56\pm0.49$ & 0.434 \\ \hline
			    & $4.59\pm0.24$ & $218_{-11}^{+12}$ & $37.13\pm0.47$ & $-28.62\pm0.44$ & -\\ 
\end{longtable}

\section{Discussions}
\subsection{Distance toward R Aqr}
Previously, distance measurements toward R Aqr had mostly been done by using three methods: (1) the kinematic method using the nebulae around the system, (2) the Period-Luminosity method, and (3) the direct distance measurement by parallax. We summarize the distance measurements toward R Aqr in Table \ref{tab:LT2} including the result of the present study.

The first kinematic distance measurement toward R Aqr was introduced by Baade (1943, 1944). He deduced a distance of 260 pc based on a constant expansion velocity for the outer nebula of between 80 and 100 km s$^{-1}$ with an expansion age of about 600 years. In addition, Solf \& Ulrich (1985) described the two nebulae as having dense equatorial ring structures with equatorial expansion velocities of 32 and 55 km s$^{-1}$ for the inner and outer nebula, respectively. By adopting the expansion age for the outer nebula of 600 years and the inner nebula of 180 years from Baade (1944) and Sopka et al. (1982) respectively, they deduced the distance of 180 pc and 185 pc for the outer and the inner nebula respectively. Recently, Korean historical records provided two irregular outbursts called \textquotedblleft Guest Star\textquotedblright epochs in A.D. 1073 and 1074 (Yang et al. 2005). Based on the records, they estimated the kinematic distance of about 273 pc with a constant expansion velocity of 55 km s$^{-1}$ from Solf \& Ulrich (1985).

In pulsating stars, one noticeable and useful property is the Period-Luminosity relation. This relation has been an important indicator of distances toward pulsating stars.

The recent Period-Luminosity relations in AGB stars was provided by Whitelock et al. (2008), who reanalyzed the revised Hipparcos parallaxes and compared them with VLBI parallax measurements. They established the Period-Luminosity relation in the infrared K-band magnitude:
\begin{equation}
M_{K}= \rho (\log P - 2.38) + \delta,
\end{equation}
where they obtained a slope of $\rho=-3.51\pm0.20$ and a zero point of $\delta=-7.25\pm0.05$ for the O-rich Mira variables in our Galaxy. Using their Period-Luminosity relation, Whitelock et al. (2008) presented a distance of 250 pc for R Aqr. 

According to Hipparcos observations (Perryman 1997), a parallax of R Aqr was $5.07\pm3.15$ mas, which corresponds to a distance of $197_{-75}^{+323}$ pc. More accurate parallax measurement was achieved by Kamohara et al. (2010) using the SiO maser distributions around Mira variable in R Aqr with VERA observations. They followed the center of circular ring fitting of the SiO maser distributions, and yielded an annual parallax of $\pi=4.7\pm0.8$ mas, corresponding to a distance of $214_{-32}^{+45}$ pc.

Our parallax measurement of $\pi = 4.59\pm0.24$ mas ($D = 218_{-11}^{+12}$ pc) is consistent with previous parallax measurement results, but yields the distance with the highest accuracy of about 5\% level. Compared with the distance measurements from other methods, our result is smaller than the distance obtained by the Period-Luminosity relation. According to the Period-Luminosity relation by Whitelock et al. (2008), expected absolute K-band magnitude for R Aqr is $M_{\mathrm{K}} = -7.98 \pm 0.09$ with the pulsation period of 387 days. Adopting an apparent K-band magnitude of $m_{\mathrm{K}} = -1.02$  obtained from SAAO (South African Astronomical Observatory) by Whitelock et al. (2000), our distance measurement provides the absolute magnitude of $M_{\mathrm{K}} = -7.71 \pm 0.11$. Although this value does not include the error in apparent K-band magnitude, our distance result presents a lower K-band absolute magnitude than that provided by the Period-Luminosity relation in Whitelock et al. (2008). On the other hand, Whitelock et al. (2008) also presented that the Period-Luminosity relation estimated by only VLBI parallaxes of 5 OH-Mira provides the zero point of $\delta = -7.08 \pm 0.17$. This relation yields the absolute K-band magnitude of $M_{\mathrm{K}} = -7.82 \pm 0.16$, which is consistent with our result. However, more VLBI parallax measurements are needed to establish a reliable Period-Luminosity relation in the future since the number of VLBI parallax measurements for Mira variables are only five in their relation. 

Moreover, kinematic distances for R Aqr have a large range from 180 to 270 pc based on the nebula properties such as an angular size, an expansion age, and an expansion velocity of the nebula. In the case of the inner nebula, kinematic properties with the equatorial shell radius of 6.5 arcsec and the equatorial expansion velocity of 32 km s$^{-1}$ were derived using a kinematic model from Solf \& Ulrich (1985). Applying our distance to their properties, the equatorial shell radius for inner nebula corresponds to the scale of $1494^{+78}_{-70}$ AU. We assumed that the inner nebula have a constant expansion rate with the velocity from Solf \& Ulrich (1985), and then estimated that the expansion age of the inner nebula is about 240 years, corresponding to about the year 1773.

On the other hand, outer nebula is suggested to have been formed in A.D. 1073 and 1074 from the outburst record found by Yang et al. (2005). Recently, a nitrate ion recorded in the Antarctic ice core also supported a nova eruption of R Aqr between A.D 1060 and 1080 (Tanabe \& Motizuki 2012). In the case of the outer nebula, kinematic properties with the equatorial shell radius of 42 arcsec and the expansion velocity of 55 km s$^{-1}$ were also reported by Solf \& Ulrich (1985). The outburst record with our distance measurement suggested that the assumed constant expansion velocity for outer nebula is about $48 \pm 2$ km s$^{-1}$ with  an angular size of 42 arcsec. This suggests that the expansion velocity of the outer nebula might be smaller than 55 km s$^{-1}$ or expansion of the outer nebula might be accelerated.

\begin{longtable}{lcc}
  \caption{Historical distance measurements toward R Aqr}\label{tab:LT2}
  \hline              
	& Distance & Method \\ \hline
 \endfirsthead
  \hline
\endhead
  \hline
\endfoot
  \hline
\endlastfoot
  \hline
			 Baade (1943)            & 260 pc                & outer nebula kinematic \\
			 Solf \& Ulrich (1985)   & 180 pc                & outer nebula kinematic \\ 
			 Solf \& Ulrich (1985)   & 185 pc                & inner nebula kinematic \\ 
			 Yang et al. (2005)      & 273 pc                & outer nebula kinematic \\
			 Whitelock et al. (2008) & 250 pc                & Period-Luminosity relation \\
			 Perryman (1997)		 & $197_{-75}^{+323}$ pc & Hipparcos parallax \\ 
 			 Kamohara et al. (2010)  & $214_{-32}^{+45}$ pc  & VERA parallax \\ 
 			 Our result              & $218_{-11}^{+12}$ pc  & VERA parallax \\ \hline
\end{longtable}


\subsection{Proper motion of the maser spot}

Along with the parallax measurement, tracing the maser spot also presents the linear proper motion of $\mu_{\alpha}^{*}=37.13 \pm 0.47$ mas yr$^{-1}$ and $\mu_{\delta}=-28.62 \pm 0.44$ mas yr$^{-1}$ in right ascension and declination, respectively. By converting to a physical scale, the proper motion of the maser spot is equivalent to about 48.45 km s$^{-1}$ in the Southeast direction (1 mas yr$^{-1}$ corresponds to 1.03 km s$^{-1}$ at the distance of 218 pc).   

According to the result from Kamohara et al. (2010), the linear proper motion of R Aqr was $\mu_{\alpha}^{*}=32.2 \pm 0.8$ mas yr$^{-1}$ and $\mu_{\delta}=-29.5 \pm 0.7$ mas yr$^{-1}$, which includes the secular motion and the binary motion of the central star. Our linear proper motion of the maser spot is also composed of the secular motion and the binary motion as well as individual motion of the maser spot, so that the subtracted proper motion vector from the result of Kamohara et al. (2010) indicates the internal individual motion of the maser spot in the maser-emitting region. The subtracted motion of the maser spot with respect to the motion of the central star is $4.93\pm1.27$ mas yr$^{-1}$ and $0.88\pm1.14$  mas yr$^{-1}$ in right ascension and declination respectively. A magnitude of the proper motion is about 5.17 km s$^{-1}$ in an eastward direction with a position angle of about 280 degrees. 

\begin{figure*}
  \begin{center}
  	\includegraphics[clip=true, width=130mm]{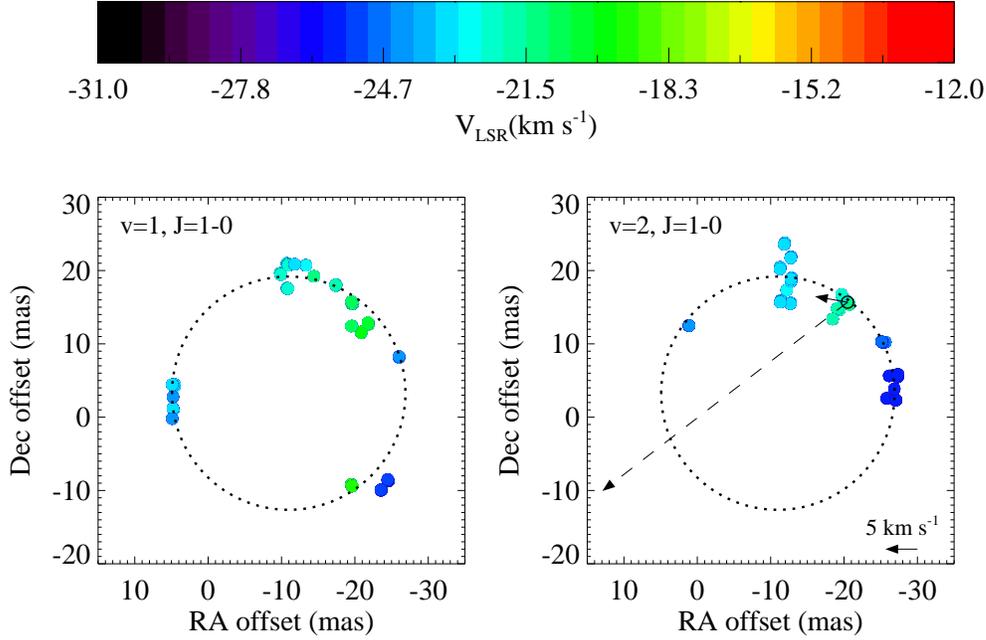}
  \end{center}
  \caption{Distribution of phase-referencing SiO maser spots toward R Aqr and fitted circle from both maser transitions at 2005/270. Left : SiO $v=1$, $J=1-0$ maser transition distribution. Right : SiO $v=2$, $J=1-0$ maser transition distribution. The center coordinate is $(\alpha, \delta)_{\mathrm{J2000}} = (\timeform{23h43m 49s.4736}, \timeform{-15D17’04".3620})$. The top color bar denoted the LSR velocity range of -12 to -31 km s$^{-1}$. The open black circle symbol indicates the maser spot for parallax measurement. The dashed arrow indicates the linear proper motion of the traced maser spot, and solid arrow indicates the subtracted proper motion from the motion of central star from Kamohara et al. (2010). The dotted circle presents the circular fitting of both SiO transitions with radius of 15.90 mas.}\label{fig:3}
\end{figure*}

Obtained distributions of the $v=1$ and $v=2$, $J=1-0$ SiO maser spots and proper motion vectors are displayed in Figure \ref{fig:3}. We also presented a result of circular fitting to the both transitions of SiO maser spots in Figure \ref{fig:3}. In the maser distributions, the traced maser spot by denoting a black circle in Figure \ref{fig:3} is likely to show a motion along the maser-emitting shell that corresponds to the circle of SiO maser distribution. Overall motion of the SiO maser shell was suggested to be contracting by Kamohara et al. (2010), who estimate the infall velocity of 3.1 $\pm$ 0.6 km s$^{-1}$. However, in our result, the proper motion of the traced maser spot does not seem to have any inward motion. Although overall motions of maser features have a radial motion of expansion or contraction following the motion of the SiO maser-emitting region in the circumstellar envelope, some cases of maser components appear to move arbitrarily due to complex dynamics of the circumstellar SiO maser shell. 

In the previous VLBI observations by Boboltz et al. (1997), they also provided an average inward motion of individual maser components with mean infall speed of 4.2 $\pm$ 0.9 km s$^{-1}$. However, some of the components appeared to move arbitrarily with respect to inward motion. Individual SiO maser components in TX Cam also presented complex motions not only in the plane of the sky, but also along the line of sight. These motions are thought to be attributable to turbulence in the maser shell, and also to changes in the conditions conducive to maser emission (Gronidakis et al. 2010). 

A rotating SiO maser shell proposed by Hollis et al. (2000, 2001) was also a proposal considered for our proper motion result. However, velocity gradients of the SiO maser distributions were inconsistent with Hollis et al. (2001), and no observation epochs were compatible with the rotating maser shell.


\bigskip

We would like to thank all the VERA staff members for their assistance concerning the array operations and data correlations. This work was supported in part by The Graduate University for Advanced Studies (Sokendai).




\end{document}